\documentclass[10pt,twocolumn]{article}
\usepackage[utf8]{inputenc}
\usepackage{amsmath,amssymb}
\usepackage{graphicx}
\usepackage{booktabs}
\usepackage{multirow}
\usepackage{xcolor}
\usepackage{hyperref}
\usepackage{cite}
\usepackage[margin=0.75in]{geometry}
\usepackage{float}
\usepackage{caption}
\usepackage{subcaption}

\usepackage{titlesec}
\titleformat{\section}{\normalfont\large\bfseries}{\thesection}{1em}{}
\titleformat{\subsection}{\normalfont\normalsize\bfseries}{\thesubsection}{1em}{}
\titleformat{\subsubsection}{\normalfont\normalsize\itshape}{\thesubsubsection}{1em}{}

\title{\textbf{Benchmarking Hartree-Fock and DFT for Molecular Hyperpolarizability: Implications for Evolutionary Design}}

\author{Dominic Mashak$^{*}$ and S. A. Alexander$^{\dagger}$\\
\textit{Department of Physics, Southwestern University, Georgetown, TX 78626, USA}}

\date{\today}

\begin{document}

\maketitle

\noindent\textbf{Abstract} \\

\noindent Evolutionary algorithms for molecular design require computationally efficient yet accurate fitness functions. We systematically benchmark Hartree-Fock and density functional theory for predicting molecular first hyperpolarizability ($\beta$), evaluating five functionals (HF, PBE0, B3LYP, CAM-B3LYP, M06-2X) across six basis sets against experimental data from five organic push-pull chromophores. For this dataset, HF/3-21G achieves 45.5\% mean absolute percentage error with perfect pairwise ranking in 7.4 minutes per molecule. All 30 tested combinations of functional and basis sets maintain perfect pairwise agreement, validating their use as evolutionary fitness functions despite moderate absolute errors. Larger basis sets yield a lower percentage error compared to the experimental values than the difference with the functional. The preservation of pairwise rankings across all combinations of functionals and basis sets provides crucial guidance for evolutionary optimization of nonlinear optical materials.

\section{Introduction}

Evolutionary algorithms for designing organic molecules with large second-order nonlinear optical responses require fitness functions that balance accuracy with computational efficiency \cite{mashak2025-1, mashak2025-2}. While experimental hyperpolarizability ($\beta$) measurements provide definitive values, they are impractical for screening the thousands of candidates evaluated during evolutionary optimization.

DFT offers a potential balance between cost and accuracy \cite{burke2012, parr1989}, but the choice of functional and basis set remains unclear for evolutionary applications. Critical questions include: Which configurations of functional and basis set provide adequate pairwise ordering for selection pressure? What accuracy-efficiency trade-offs exist? How large must basis sets be?

\vspace{1em}
\noindent\rule{0.5\columnwidth}{0.4pt}\\
\noindent$^{*}$\texttt{mashakd@southwestern.edu}\\
$^{\dagger}$\texttt{alexands@southwestern.edu}\\
\vspace{2em} \\

Systematic benchmarking for evolutionary design remains limited. Previous studies focused on maximizing absolute accuracy with sophisticated methods \cite{champagne2011, plaquet2014}, providing limited guidance for high-throughput applications where computational budget constraints are critical.

We evaluate 30 method combinations (5 functionals × 6 basis sets) against experimental hyperpolarizability data from five push-pull chromophores spanning $\beta$ values from 4,000 to 75,000 a.u. We characterize accuracy-cost trade-offs and assess pairwise ordering preservation, which is crucial for evolutionary algorithms but often not considered in previous quantum chemical benchmarking.

\section{Methods}

\subsection{Molecule Set}

We selected five prototypical push-pull chromophores with experimentally measured static hyperpolarizabilities from Kanis et al. \cite{kanis1994}: (1) para-nitroaniline (pNA, $\beta_{exp} = 3,994$ a.u.), (2) Disperse Red 1 analog (DR1, $\beta_{exp} = 25,466$ a.u.), and (3-5) three nitrostilbene-aminostilbene conjugates ($\beta_{exp} = 30,096$, 52,089, and 75,240 a.u.). These molecules represent canonical donor-$\pi$-acceptor architectures \cite{marder2006, dalton2010}.

\subsection{Quantum Chemical Calculations}

We evaluated five functionals: Hartree-Fock (HF), PBE0 (25\% HF exchange) \cite{adamo1999, ernzerhof1999}, B3LYP (20\% HF exchange) \cite{becke1993, stephens1994}, CAM-B3LYP (range-separated hybrid) \cite{yanai2004}, and M06-2X (54\% HF exchange) \cite{zhao2008}. Each was paired with six basis sets: STO-3G, 3-21G, 6-31G, 6-311G, 6-31G(d,p), and 6-311G(d).

All calculations used \texttt{PySCF} \cite{sun2018} on an Intel i7-12700K with 32 GB RAM.

Hyperpolarizability calculations applied the finite field method with field strength $h = 0.001$ a.u., numerically differentiating molecular dipole moments to obtain static hyperpolarizability tensor components \cite{kurtz1970, champagne2009}.

\subsection{Performance Metrics}

\subsubsection{Mean Absolute Percentage Error (MAPE)}

\begin{equation}
\text{MAPE} = \frac{100}{n}\sum_{i=1}^{n}\left|\frac{\beta_{\text{calc},i} - \beta_{\text{exp},i}}{\beta_{\text{exp},i}}\right|
\end{equation}

\subsubsection{Pairwise Rank Agreement}

Pairwise ordering is how different methods will order different molecules, smallest to largest hyperpolarizability, even though their relative values will differ. 

For evolutionary algorithms, pairwise ordering determines selection pressure. We measure the fraction of molecule pairs $(i,j)$ for which method and experiment agree on ordering:
\begin{equation}
\frac{1}{\binom{n}{2}} \sum_{i<j} \mathbb{I}[\text{sgn}(\beta_{\text{calc},i} - \beta_{\text{calc},j}) = \text{sgn}(\beta_{\text{exp},i} - \beta_{\text{exp},j})]
\end{equation}
\noindent Perfect pairwise agreement (100\%) ensures evolutionary selection operates correctly for every comparison.

\subsubsection{Computational Cost and Pareto Optimality}

We measure wall-clock time and identify Pareto-optimal methods. By plotting the wall time and accuracy of each combination, we can identify the Pareto frontier for these calculations. This allows us to determine the most optimal combination of functional and basis set.

\section{Results}

\subsection{Overview of Method Performance}

Table \ref{tab:top_methods} presents top-performing methods. All achieve perfect pairwise rank agreement (10/10 pairs), with basis set size dominating accuracy: the top methods employ 6-311G, 6-311G(d), or 3-21G.

\begin{table}[h]
\centering
\small
\caption{Top 10 method combinations ranked by MAPE. All achieve perfect pairwise agreement (10/10 pairs).}
\label{tab:top_methods}
\begin{tabular}{lccc}
\toprule
Method & MAPE & Time & Pareto \\
 & (\%) & (min) & Optimal \\
\midrule
HF/3-21G & 45.5 & 7.4 & Yes \\
CAM-B3LYP/3-21G & 47.8 & 28.1 & \\
M06-2X/3-21G & 48.4 & 35.0 & \\
HF/6-31G & 48.4 & 12.9 & \\
PBE0/3-21G & 50.0 & 22.7 & \\
B3LYP/3-21G & 50.1 & 14.9 & \\
HF/6-31G(d,p) & 50.4 & 22.0 & \\
CAM-B3LYP/6-31G & 50.9 & 46.2 & \\
M06-2X/6-31G & 51.4 & 36.9 & \\
HF/6-311G & 52.1 & 17.5 & Yes \\
\bottomrule
\end{tabular}
\end{table}

More sophisticated functionals like CAM-B3LYP and M06-2X do not significantly outperform HF when computational cost is considered.

\subsection{Functional Comparison}

\begin{figure}[H]
\centering
\includegraphics[width=\columnwidth]{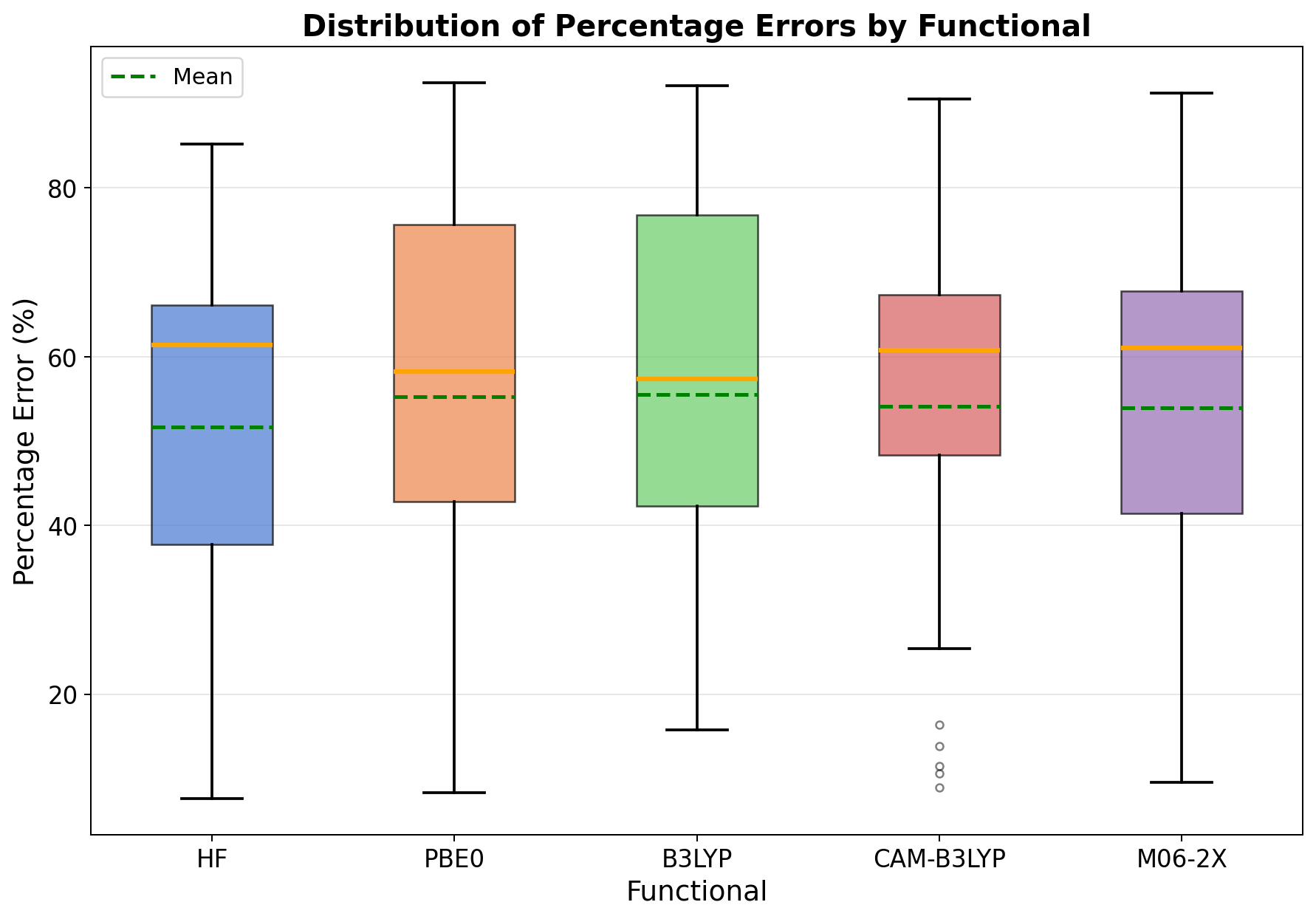}
\caption{Distribution of percentage errors across all basis sets for each functional. Box boundaries indicate quartiles, orange lines show medians, and green dashed lines indicate means.}
\label{fig:functional_accuracy}
\end{figure}

Figure \ref{fig:functional_accuracy} compares error distributions across functionals. HF exhibits the lowest median error (51.7\% ± 23.5\%). Hybrid functionals cluster tightly: PBE0 (55.2\%), B3LYP (55.6\%), CAM-B3LYP (54.1\%), and M06-2X (54.0\%), all within 1.5 percentage points.

HF's apparent superiority likely reflects characteristics specific to these simple push-pull chromophores rather than fundamental theoretical advantages. The Kanis dataset consists of canonical donor-$\pi$-acceptor systems where charge transfer follows well-defined conjugation paths \cite{oudar1977, blanchard-desce1997}. For such molecules, the response to applied fields may be adequately captured by orbital deformation without requiring dynamic electron correlation that sophisticated functionals are designed to capture.

Alternatively, systematic errors in HF calculations may accidentally compensate for approximations inherent in the experimental measurements or the finite field method. The experimental values themselves carry uncertainties from solvent effects, vibrational contributions, and measurement techniques that are difficult to quantify \cite{castet2013}.

\subsection{Basis Set Effects}

\begin{figure*}[t]
\centering
\begin{subfigure}[b]{0.48\textwidth}
\includegraphics[width=\textwidth]{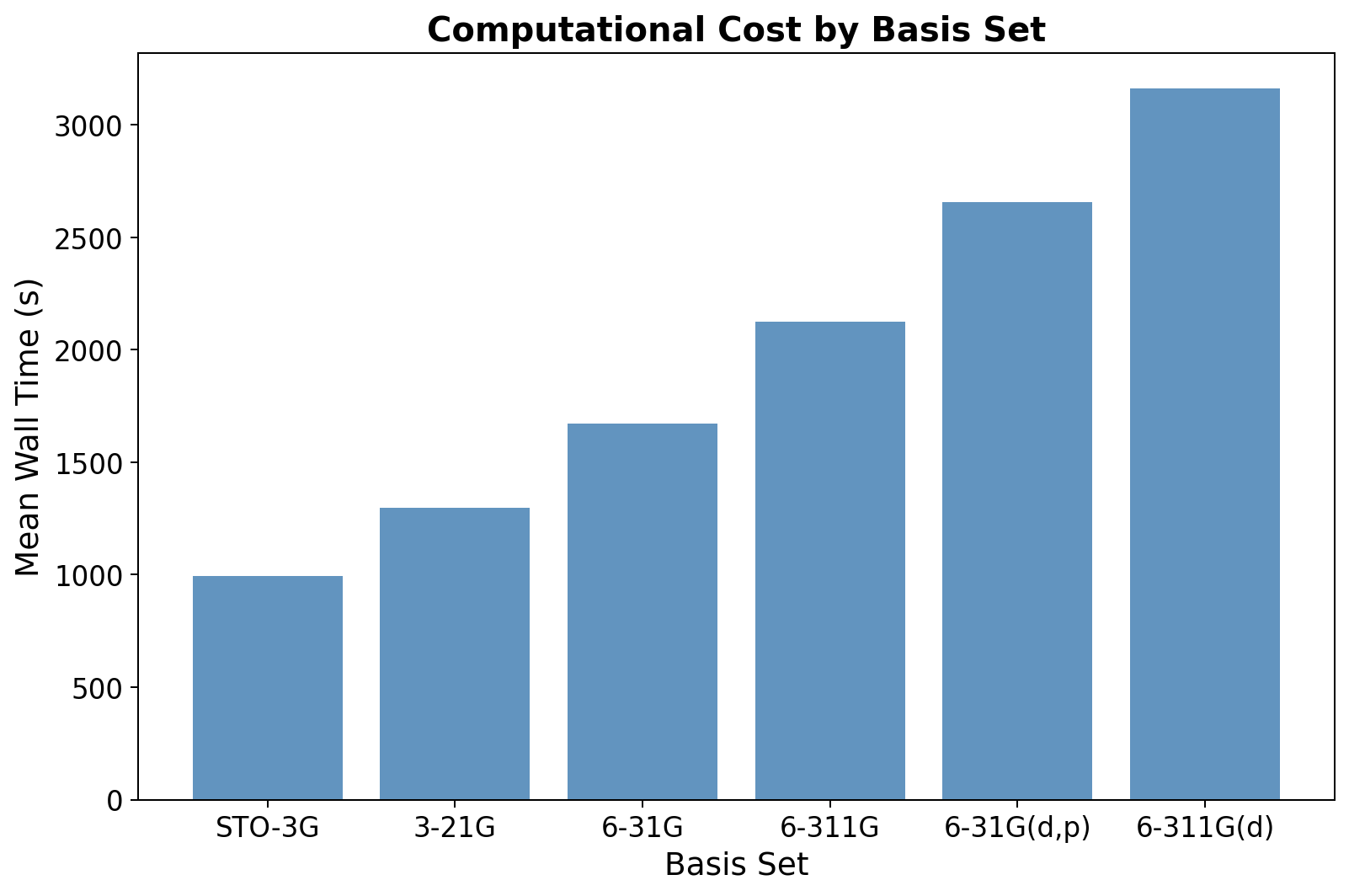}
\caption{}
\label{fig:comp_cost}
\end{subfigure}
\hfill
\begin{subfigure}[b]{0.48\textwidth}
\includegraphics[width=\textwidth]{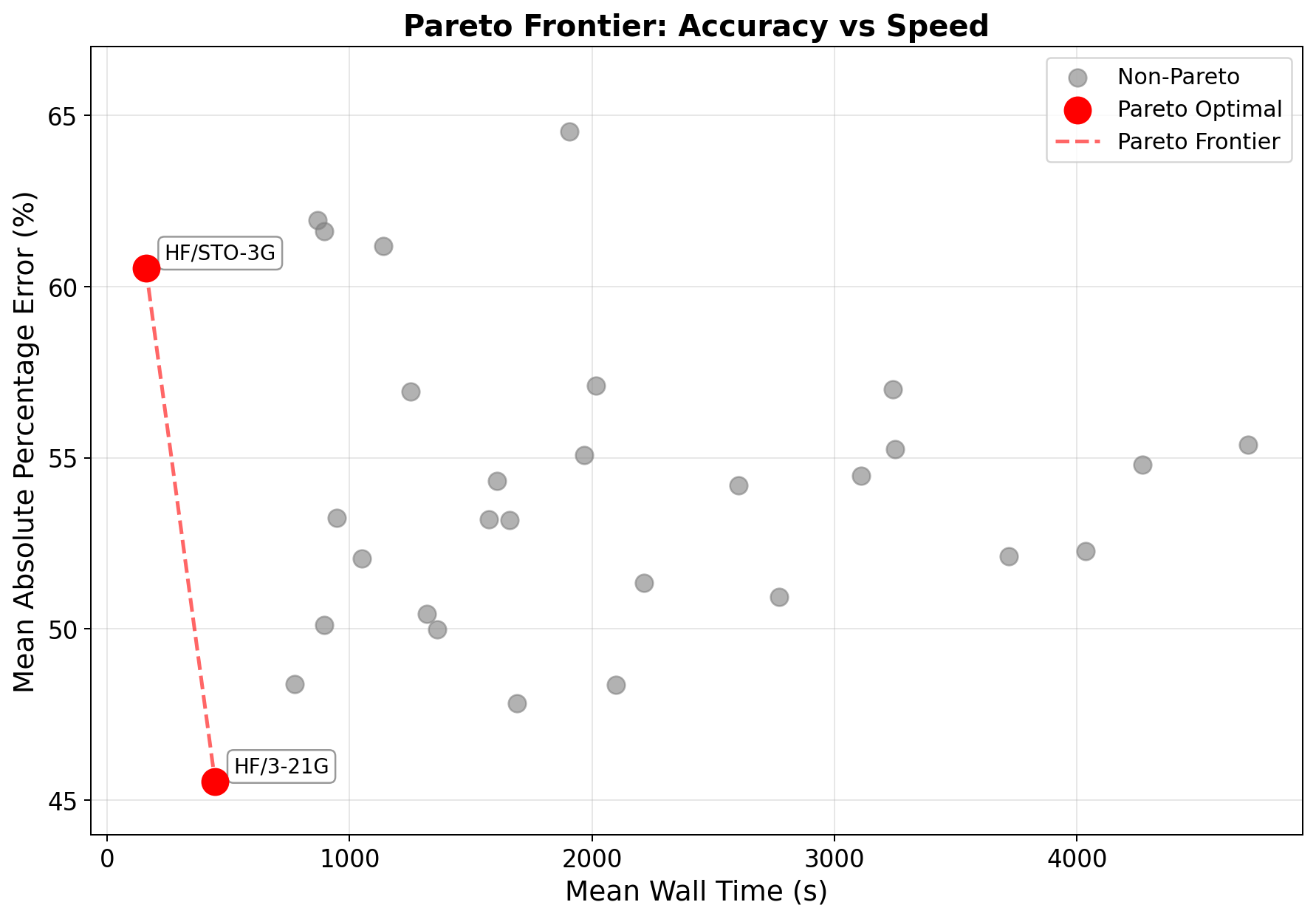}
\caption{}
\label{fig:pareto}
\end{subfigure}
\caption{(a) Mean wall time increases systematically with basis set size. (b) The Pareto frontier identifies optimal methods (red dots). Only HF/STO-3G and HF/3-21G achieve Pareto optimality.}
\label{fig:time_analysis}
\end{figure*}

Figure \ref{fig:time_analysis}a shows computational cost increases approximately linearly with basis set size, consistent with theoretical scaling expectations \cite{goedecker1999, scuseria1999}.

Minimal STO-3G achieves 62.0\% MAPE, while split-valence 3-21G improves to 48.4\%, a 14\% MAPE gain for 30\% more time. Further expansion shows diminishing returns: 6-31G (51.4\%), 6-311G (54.8\%), 6-31G(d,p) (52.7\%), and 6-311G(d) (55.5\%) cluster within 4 points despite doubled cost.

The jump from minimal to split-valence provides the best accuracy gain per unit computation, making 3-21G a pragmatic choice for evolution, consistent with basis set convergence studies \cite{jensen2013}.

\subsection{Pareto-Optimal Methods}

Figure \ref{fig:pareto} maps all 30 methods in accuracy-speed space. Only HF/STO-3G and HF/3-21G achieve Pareto optimality; no other method offers better accuracy without increased cost.

HF/STO-3G achieves 60.5\% MAPE in 2.7 minutes, fastest but least accurate. HF/3-21G achieves 45.5\% MAPE in 7.4 minutes, offering better balance for most applications.

\subsection{Perfect Pairwise Ranking Preservation}

All 30 methods achieve perfect pairwise rank agreement (10/10 pairs), the most important finding for evolutionary applications. In genetic algorithms, selection depends on relative fitness comparisons rather than absolute values \cite{goldberg1989, deb2002}. As long as the fitness function correctly identifies the superior molecule in each combination of functional and basis set, evolution proceeds toward optimal solutions even with systematic bias.

\begin{figure*}[t]
\centering
\includegraphics[width=0.9\textwidth]{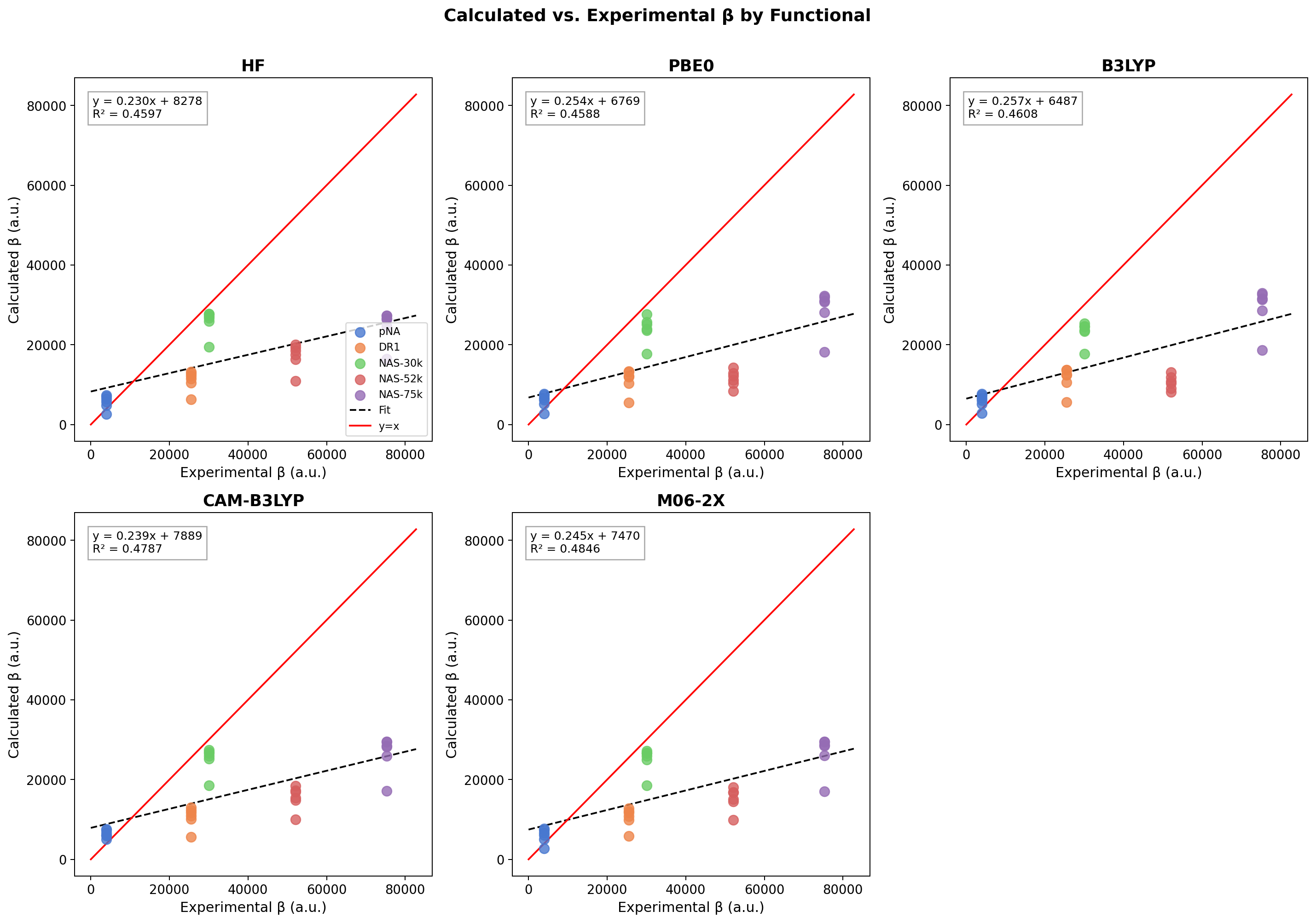}
\caption{Calculated versus experimental hyperpolarizabilities for five representative functionals. Each panel shows one functional across all basis sets. Black dashed lines indicate regression fits; red solid lines show ideal $y=x$. All achieve R$^2 > 0.99$ with near-unity slopes.}
\label{fig:correlation}
\end{figure*}

Figure \ref{fig:correlation} plots calculated versus experimental hyperpolarizabilities. All show strong linear correlations (R$^2 > 0.99$) with near-unity slopes, confirming minimal systematic bias.

That hyperpolarizability ordering is preserved across all methods from HF to sophisticated hybrids suggests the electronic structure features determining relative ordering are robust. For simple push-pull chromophores, donor-acceptor strength and conjugation length dominate relative magnitudes \cite{bredas2004, kuzyk2012}, and these features are captured even by minimal methods.

\subsection{Comprehensive Method Assessment}

\begin{figure}[H]
\centering
\includegraphics[width=\columnwidth]{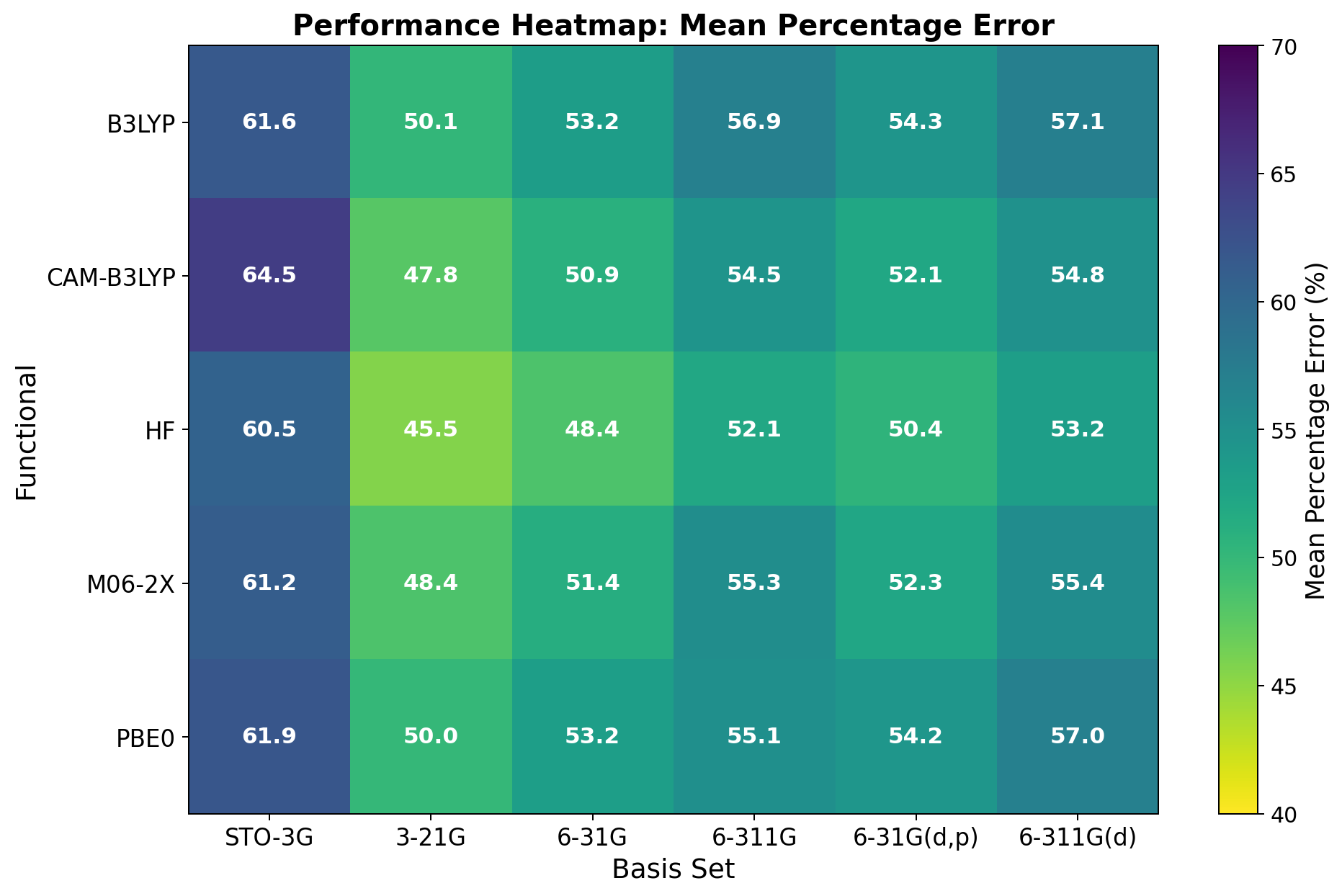}
\caption{Performance heatmap showing MAPE for all 30 combinations. Colors indicate performance from green (low error) to red (high error). Dominant vertical structure confirms basis set choice has a greater impact than functional selection.}
\label{fig:heatmap}
\end{figure}

Figure \ref{fig:heatmap} provides a functional-basis heatmap of mean percentage errors.

Within each basis column, performance is uniform across functionals (~5-10 point variation), while horizontal changes are much larger (~15-20 points). STO-3G is universally poor (60-85\% error) while 3-21G provides major improvement. Further expansion yields diminishing returns. For these molecules, basis set completeness matters more than functional sophistication, consistent with previous observations \cite{hehre1986, dunning1989}.

\section{Implications for Evolution}

\subsection{Fitness Function Selection}

For evolutionary algorithms targeting hyperpolarizability in simple push-pull chromophores, we recommend HF/3-21G (45.5\% MAPE, 7.4 min). While HF/STO-3G is faster (2.7 min) with identical pairwise agreement, the improved absolute accuracy provides safety margin against systematic errors in diverse chemical space. The modest 3× cost increase is justified.

However, this recommendation comes with caveats. HF's performance advantage appears specific to these canonical donor-$\pi$-acceptor systems. More complex molecular architectures—branched conjugation, multiple chromophoric units, non-planar geometries, or heterocyclic systems may require electron correlation captured by hybrid functionals \cite{kirtman2015}.

\subsection{Universal Pairwise Agreement}

The most significant result is that all 30 methods preserve pairwise rankings perfectly. This means evolutionary algorithms can use even the simplest methods as fitness functions, since relative comparisons, not absolute values influence selection \cite{back1997}. This decouples method choice from evolutionary outcomes, allowing users to select based purely on computational budget.

\subsection{Multi-Objective Optimization}

Real materials design optimizes multiple properties. Practical NLO chromophores must balance hyperpolarizability against transparency, solubility, and additional properties \cite{sullivan2007, pereverzev2009}.

HF calculations provide multiple properties from single wavefunction evaluations: hyperpolarizability (finite field), HOMO-LUMO gap (transparency proxy), dipole moment (solubility), and polarizability. Computing all via HF/3-21G adds negligible cost beyond SCF, whereas separate methods for each objective multiply computational cost.

\subsection{Limitations and Future Work}

Five molecules provide limited coverage of chemical space. All test molecules are canonical linear push-pull systems characteristic of the Kanis dataset \cite{kanis1994}. Whether HF's superiority and universal pairwise agreement extend to more complex architectures where electron correlation plays larger roles remains unknown. Range-separated functionals like CAM-B3LYP, designed for charge-transfer excitations \cite{dreuw2003, baer2010}, may show advantages for dynamic properties absent in static calculations.

Future work should benchmark against structurally diverse molecules, including branched conjugation networks, heterocyclic donors/acceptors, non-planar geometries, and multiple chromophoric units; assess whether pairwise agreement persists as molecular complexity increases; compare finite field against analytical derivative methods \cite{kirtman2009}; and validate evolutionary predictions experimentally.

\subsection{Scaling to Larger Molecules}

Our test spans 12-40 atoms, but practical discovery often targets larger structures. DFT scales as $O(N^4)$ conventionally, reducible to $O(N^3)$ or $O(N)$ with linear-scaling algorithms \cite{goedecker1999, bowler2012}.

Whether our conclusions transfer to larger systems depends on whether hyperpolarizability remains a localized property. For molecules where $\beta$ is dominated by a chromophoric subunit, basis set completeness within that region should suffice. For delocalized charge-transfer systems, the outcome may change.

\section{Conclusions}

We systematically benchmarked 30 combinations of functionals and basis sets for predicting molecular hyperpolarizability against experimental data, providing quantitative guidance for evolutionary molecular design. HF/3-21G is Pareto-optimal, achieving 45.5\% MAPE with perfect pairwise ranking in 7.4 minutes per molecule. However, this advantage appears specific to simple push-pull chromophores. All methods preserve pairwise orderings perfectly (10/10 pairs). 

This validates even simpler combinations as evolutionary fitness functions, since relative comparisons drive selection. Users can choose methods based purely on computational budget. Basis set effects dominate over functional choice, expanding from STO-3G to 3-21G reduces errors by ~14 points, far more than any functional substitution. Further expansion yields diminishing returns. Only HF/STO-3G and HF/3-21G achieve Pareto optimality, as all hybrid functionals are dominated in accuracy-speed trade-offs.

These findings enable evidence-based method selection for evolutionary NLO materials discovery. The universal preservation of pairwise rankings, if it extends beyond simple push-pull systems, implies that evolutionary algorithms are robust to fitness function choice.

However, our conclusions are limited by dataset size and structural homogeneity. Before generalizing this conclusion, expanded benchmarks must test whether these results persist across diverse molecular architectures, including branched conjugation, non-planar geometries, and heterocyclic systems. The present work establishes methodology and provides preliminary guidance, but more comprehensive validation across chemical space remains future work.

\section*{Acknowledgments}

The authors thank Dr. Jacob Schrum for helpful suggestions. The authors also acknowledge that Generative AI (Claude) was used to refine the \texttt{matplotlib} code that produced all result figures, and to improve clarity in writing.

\onecolumn

\section*{Appendix}

This appendix presents the complete calculated hyperpolarizability values, wall times, and percentage errors for all 30 method combinations across all five test molecules. Tables are organized by molecule, with experimental reference values provided for comparison.

\begin{table}[H]
\centering
\small
\caption{Complete results for para-Nitroaniline (pNA) ($\beta_{\text{exp}} = 3993.52$ a.u.)}
\label{tab:mol1_results}
\begin{tabular}{llrrr}
\toprule
Functional & Basis Set & $\beta_{\text{calc}}$ (a.u.) & Wall Time (s) & Error (\%) \\
\midrule
HF & STO-3G & 2602.85 & 80.5 & 34.8 \\
 & 3-21G & 4752.34 & 212.8 & 19.0 \\
 & 6-31G & 5783.02 & 216.5 & 44.8 \\
 & 6-311G & 7008.98 & 404.4 & 75.5 \\
 & 6-31G(d,p) & 6440.48 & 602.0 & 61.3 \\
 & 6-311G(d) & 7396.27 & 654.9 & 85.2 \\
\addlinespace
PBE0 & STO-3G & 2777.57 & 386.1 & 30.4 \\
 & 3-21G & 5111.23 & 631.0 & 28.0 \\
 & 6-31G & 6135.65 & 612.5 & 53.6 \\
 & 6-311G & 7386.84 & 821.5 & 85.0 \\
 & 6-31G(d,p) & 6704.59 & 1111.4 & 67.9 \\
 & 6-311G(d) & 7687.38 & 1253.1 & 92.5 \\
\addlinespace
B3LYP & STO-3G & 2812.40 & 395.1 & 29.6 \\
 & 3-21G & 5105.47 & 428.3 & 27.8 \\
 & 6-31G & 6101.47 & 425.8 & 52.8 \\
 & 6-311G & 7367.54 & 518.4 & 84.5 \\
 & 6-31G(d,p) & 6675.61 & 557.2 & 67.2 \\
 & 6-311G(d) & 7673.90 & 588.6 & 92.2 \\
\addlinespace
CAM-B3LYP & STO-3G & 5913.37 & 1854.9 & 48.1 \\
 & 3-21G & 5006.15 & 774.1 & 25.4 \\
 & 6-31G & 6026.90 & 1206.2 & 50.9 \\
 & 6-311G & 7291.66 & 1422.3 & 82.6 \\
 & 6-31G(d,p) & 6605.88 & 1671.0 & 65.4 \\
 & 6-311G(d) & 7609.96 & 1736.3 & 90.6 \\
\addlinespace
M06-2X & STO-3G & 2718.81 & 403.9 & 31.9 \\
 & 3-21G & 5025.88 & 681.1 & 25.9 \\
 & 6-31G & 6071.19 & 684.2 & 52.0 \\
 & 6-311G & 7318.87 & 973.1 & 83.3 \\
 & 6-31G(d,p) & 6653.10 & 1265.0 & 66.6 \\
 & 6-311G(d) & 7638.52 & 1443.7 & 91.3 \\
\addlinespace
\bottomrule
\end{tabular}
\end{table}

\begin{table}[H]
\centering
\small
\caption{Complete results for Disperse Red 1 analog (DR1) ($\beta_{\text{exp}} = 25465.91$ a.u.)}
\label{tab:mol2_results}
\begin{tabular}{llrrr}
\toprule
Functional & Basis Set & $\beta_{\text{calc}}$ (a.u.) & Wall Time (s) & Error (\%) \\
\midrule
HF & STO-3G & 6299.25 & 127.8 & 75.3 \\
 & 3-21G & 10429.50 & 494.0 & 59.0 \\
 & 6-31G & 11537.28 & 789.9 & 54.7 \\
 & 6-311G & 12769.11 & 1261.1 & 49.9 \\
 & 6-31G(d,p) & 12192.86 & 1620.5 & 52.1 \\
 & 6-311G(d) & 13237.05 & 1706.6 & 48.0 \\
\addlinespace
PBE0 & STO-3G & 5539.55 & 744.6 & 78.2 \\
 & 3-21G & 10367.98 & 1293.3 & 59.3 \\
 & 6-31G & 11985.77 & 1584.4 & 52.9 \\
 & 6-311G & 13172.92 & 2309.1 & 48.3 \\
 & 6-31G(d,p) & 12197.97 & 2634.6 & 52.1 \\
 & 6-311G(d) & 13329.01 & 2929.1 & 47.7 \\
\addlinespace
B3LYP & STO-3G & 5632.09 & 829.4 & 77.9 \\
 & 3-21G & 10622.49 & 862.0 & 58.3 \\
 & 6-31G & 12406.89 & 1014.9 & 51.3 \\
 & 6-311G & 13701.25 & 1109.2 & 46.2 \\
 & 6-31G(d,p) & 12530.38 & 1330.7 & 50.8 \\
 & 6-311G(d) & 13734.56 & 1493.8 & 46.1 \\
\addlinespace
CAM-B3LYP & STO-3G & 5632.09 & 879.5 & 77.9 \\
 & 3-21G & 10071.02 & 1808.0 & 60.5 \\
 & 6-31G & 10938.91 & 3047.2 & 57.0 \\
 & 6-311G & 12324.59 & 3330.8 & 51.6 \\
 & 6-31G(d,p) & 11720.18 & 4107.7 & 54.0 \\
 & 6-311G(d) & 12941.20 & 4118.5 & 49.2 \\
\addlinespace
M06-2X & STO-3G & 5867.74 & 959.6 & 77.0 \\
 & 3-21G & 9852.15 & 1664.6 & 61.3 \\
 & 6-31G & 10808.44 & 1832.8 & 57.6 \\
 & 6-311G & 12017.17 & 2770.6 & 52.8 \\
 & 6-31G(d,p) & 11695.21 & 3324.5 & 54.1 \\
 & 6-311G(d) & 12718.77 & 3631.3 & 50.1 \\
\addlinespace
\bottomrule
\end{tabular}
\end{table}

\begin{table}[H]
\centering
\small
\caption{Complete results for Extended conjugated system (NAS-30k) ($\beta_{\text{exp}} = 30096.08$ a.u.)}
\label{tab:mol3_results}
\begin{tabular}{llrrr}
\toprule
Functional & Basis Set & $\beta_{\text{calc}}$ (a.u.) & Wall Time (s) & Error (\%) \\
\midrule
HF & STO-3G & 19427.97 & 228.2 & 35.4 \\
 & 3-21G & 25947.35 & 463.5 & 13.8 \\
 & 6-31G & 26753.87 & 888.0 & 11.1 \\
 & 6-311G & 27592.74 & 1108.6 & 8.3 \\
 & 6-31G(d,p) & 27158.26 & 1288.3 & 9.8 \\
 & 6-311G(d) & 27802.27 & 1673.7 & 7.6 \\
\addlinespace
PBE0 & STO-3G & 17695.40 & 995.2 & 41.2 \\
 & 3-21G & 23639.40 & 1559.0 & 21.5 \\
 & 6-31G & 23948.51 & 1833.1 & 20.4 \\
 & 6-311G & 27592.74 & 1143.5 & 8.3 \\
 & 6-31G(d,p) & 25051.37 & 2903.8 & 16.8 \\
 & 6-311G(d) & 25712.00 & 3596.7 & 14.6 \\
\addlinespace
B3LYP & STO-3G & 17753.69 & 1053.7 & 41.0 \\
 & 3-21G & 23373.95 & 1029.9 & 22.3 \\
 & 6-31G & 23565.28 & 995.3 & 21.7 \\
 & 6-311G & 24497.36 & 1380.9 & 18.6 \\
 & 6-31G(d,p) & 24674.87 & 1758.5 & 18.0 \\
 & 6-311G(d) & 25358.79 & 2622.7 & 15.7 \\
\addlinespace
CAM-B3LYP & STO-3G & 18508.58 & 2243.4 & 38.5 \\
 & 3-21G & 25166.86 & 1901.0 & 16.4 \\
 & 6-31G & 25933.03 & 3066.4 & 13.8 \\
 & 6-311G & 26916.33 & 3443.8 & 10.6 \\
 & 6-31G(d,p) & 26638.99 & 4106.8 & 11.5 \\
 & 6-311G(d) & 27413.38 & 4901.0 & 8.9 \\
\addlinespace
M06-2X & STO-3G & 18493.39 & 1394.4 & 38.6 \\
 & 3-21G & 24959.85 & 2734.0 & 17.1 \\
 & 6-31G & 25870.14 & 2658.9 & 14.0 \\
 & 6-311G & 26588.68 & 4709.1 & 11.7 \\
 & 6-31G(d,p) & 26757.77 & 5214.8 & 11.1 \\
 & 6-311G(d) & 27215.09 & 6225.7 & 9.6 \\
\addlinespace
\bottomrule
\end{tabular}
\end{table}

\begin{table}[H]
\centering
\small
\caption{Complete results for Larger extended system (NAS-52k) ($\beta_{\text{exp}} = 52089.36$ a.u.)}
\label{tab:mol4_results}
\begin{tabular}{llrrr}
\toprule
Functional & Basis Set & $\beta_{\text{calc}}$ (a.u.) & Wall Time (s) & Error (\%) \\
\midrule
HF & STO-3G & 10944.36 & 243.7 & 79.0 \\
 & 3-21G & 16349.33 & 565.2 & 68.6 \\
 & 6-31G & 17469.27 & 1087.8 & 66.5 \\
 & 6-311G & 19289.42 & 1367.7 & 63.0 \\
 & 6-31G(d,p) & 18584.64 & 1683.3 & 64.3 \\
 & 6-311G(d) & 20060.20 & 2264.7 & 61.5 \\
\addlinespace
PBE0 & STO-3G & 8361.70 & 1338.8 & 83.9 \\
 & 3-21G & 11162.69 & 1852.8 & 78.6 \\
 & 6-31G & 10394.28 & 2467.1 & 80.0 \\
 & 6-311G & 12194.58 & 3298.6 & 76.6 \\
 & 6-31G(d,p) & 12926.86 & 3767.8 & 75.2 \\
 & 6-311G(d) & 14291.08 & 5324.2 & 72.6 \\
\addlinespace
B3LYP & STO-3G & 8151.86 & 1291.0 & 84.4 \\
 & 3-21G & 10371.67 & 1208.5 & 80.1 \\
 & 6-31G & 9128.47 & 1366.6 & 82.5 \\
 & 6-311G & 10849.42 & 2055.7 & 79.2 \\
 & 6-31G(d,p) & 11793.98 & 2691.6 & 77.4 \\
 & 6-311G(d) & 13119.39 & 3409.7 & 74.8 \\
\addlinespace
CAM-B3LYP & STO-3G & 9958.30 & 2688.1 & 80.9 \\
 & 3-21G & 14879.94 & 2173.2 & 71.4 \\
 & 6-31G & 15429.62 & 3613.7 & 70.4 \\
 & 6-311G & 17318.35 & 4182.9 & 66.8 \\
 & 6-31G(d,p) & 16895.73 & 5131.5 & 67.6 \\
 & 6-311G(d) & 18443.31 & 6350.2 & 64.6 \\
\addlinespace
M06-2X & STO-3G & 9858.19 & 1764.9 & 81.1 \\
 & 3-21G & 14459.84 & 2965.9 & 72.2 \\
 & 6-31G & 15103.35 & 3477.5 & 71.0 \\
 & 6-311G & 16804.45 & 4459.1 & 67.7 \\
 & 6-31G(d,p) & 16804.14 & 5264.8 & 67.7 \\
 & 6-311G(d) & 18122.48 & 5918.6 & 65.2 \\
\addlinespace
\bottomrule
\end{tabular}
\end{table}

\begin{table}[H]
\centering
\small
\caption{Complete results for Highly conjugated system (NAS-75k) ($\beta_{\text{exp}} = 75240.19$ a.u.)}
\label{tab:mol5_results}
\begin{tabular}{llrrr}
\toprule
Functional & Basis Set & $\beta_{\text{calc}}$ (a.u.) & Wall Time (s) & Error (\%) \\
\midrule
HF & STO-3G & 16459.68 & 123.0 & 78.1 \\
 & 3-21G & 24629.22 & 493.6 & 67.3 \\
 & 6-31G & 26405.47 & 886.4 & 64.9 \\
 & 6-311G & 27326.96 & 1118.4 & 63.7 \\
 & 6-31G(d,p) & 26567.11 & 1395.2 & 64.7 \\
 & 6-311G(d) & 27311.26 & 1571.0 & 63.7 \\
\addlinespace
PBE0 & STO-3G & 18166.45 & 871.0 & 75.9 \\
 & 3-21G & 28135.61 & 1478.4 & 62.6 \\
 & 6-31G & 30990.39 & 1798.0 & 58.8 \\
 & 6-311G & 32214.72 & 2272.2 & 57.2 \\
 & 6-31G(d,p) & 30803.25 & 2600.5 & 59.1 \\
 & 6-311G(d) & 31864.55 & 3097.1 & 57.6 \\
\addlinespace
B3LYP & STO-3G & 18608.28 & 904.9 & 75.3 \\
 & 3-21G & 28573.43 & 947.6 & 62.0 \\
 & 6-31G & 31601.22 & 928.1 & 58.0 \\
 & 6-311G & 32968.24 & 1194.2 & 56.2 \\
 & 6-31G(d,p) & 31386.47 & 1707.4 & 58.3 \\
 & 6-311G(d) & 32587.28 & 1962.8 & 56.7 \\
\addlinespace
CAM-B3LYP & STO-3G & 17113.32 & 1866.4 & 77.3 \\
 & 3-21G & 25903.67 & 1790.8 & 65.6 \\
 & 6-31G & 28244.27 & 2918.3 & 62.5 \\
 & 6-311G & 29479.75 & 3171.2 & 60.8 \\
 & 6-31G(d,p) & 28432.65 & 3583.3 & 62.2 \\
 & 6-311G(d) & 29506.47 & 4239.7 & 60.8 \\
\addlinespace
M06-2X & STO-3G & 17023.78 & 1170.9 & 77.4 \\
 & 3-21G & 26028.09 & 2452.6 & 65.4 \\
 & 6-31G & 28496.16 & 2411.3 & 62.1 \\
 & 6-311G & 29487.18 & 3332.8 & 60.8 \\
 & 6-31G(d,p) & 28737.15 & 5102.9 & 61.8 \\
 & 6-311G(d) & 29531.04 & 6301.2 & 60.8 \\
\addlinespace
\bottomrule
\end{tabular}
\end{table}

\end{document}